\documentclass[sn-mathphys,Numbered]{sn-jnl}

\usepackage{graphicx}%
\usepackage{multirow}%
\usepackage{amsmath,amssymb,amsfonts}%
\usepackage{amsthm}%
\usepackage{mathrsfs}%
\usepackage[title]{appendix}%
\usepackage{xcolor}%
\usepackage{textcomp}%
\usepackage{manyfoot}%
\usepackage{booktabs}%
\usepackage{algorithm}%
\usepackage{algorithmicx}%
\usepackage{algpseudocode}%
\usepackage{listings}%
\usepackage{upgreek}
\usepackage{99_IMS_constants}
\usepackage{siunitx}
\usepackage{blindtext}
\usepackage[rightcaption]{sidecap}

\begin{document}

\title[Design considerations for the optimization of {\lSQ s}]{Design considerations for the optimization of {\lSQ s}}

\author*[1]{\fnm{Constantin} \sur{Schuster}}\email{constantin.schuster@kit.edu}
\author[1,2]{\fnm{Sebastian} \sur{Kempf}}

\affil[1]{\orgdiv{Institute of Micro- and Nanoelectronic Systems (IMS)}, \orgname{Karlsruhe Institute of Technology (KIT)}, \orgaddress{\street{Hertzstrasse 16}, \city{Karlsruhe}, \postcode{76187}, \country{Germany}}}
\affil[2]{\orgdiv{Institute for Data Processing and Electronics (IPE)}, \orgname{Karlsruhe Institute of Technology (KIT)}, \orgaddress{\street{Hermann-von-Helmholtz-Platz 1}, \city{Karlsruhe}, \postcode{76344}, \country{Germany}}}

\abstract{Cryogenic microcalorimeters are key tools for high-resolution X-ray spectroscopy due to their excellent energy resolution and quantum efficiency close to \SI{100}{\percent}. Multiple types of microcalorimeters exist, some of which have already proven outstanding performance. Nevertheless, they can't yet compete with cutting-edge grating or crystal spectrometers. For this reason, novel microcalorimeter concepts are continuously developed. One such concept is based on the strong temperature dependence of the magnetic penetration depth of a superconductor operated close to its transition temperature. This so-called $\lambda$-SQUID provides an in-situ tunable gain and promises to reach sub-eV energy resolution. Here, we present some design considerations with respect to the optimization of such a detector that are derived by analytic means. We particularly show that for this detector concept the heat capacity of the sensor should match the heat capacity of the absorber.
}


\maketitle

\section{Introduction}

Cryogenic microcalorimeters such as superconducting transition-edge sensors (TESs) \cite{Irwin2005,Ullom2015} or metallic magnetic calorimeters (MMCs) \cite{Fleischmann2005,Kempf2018} have proven to be outstanding tools for measuring the energy of X-ray photons with unprecedented precision. They rely on sensing the change in temperature of an X-ray absorber upon photon absorption using an extremely sensitive thermometer that is based either on a superconducting (TES) or paramagnetic (MMC) sensor material. Due to their unique combination of excellent energy resolution and quantum efficiency close to \SI{100}{\percent}, they offer significant advantages as compared to state-of-the-art grating or crystal X-ray spectrometers \cite{Uhlig2015}. Their quantum efficiency significantly relaxes the requirements on X-ray beam intensity, which especially benefits measurements on strongly diluted or radiation sensitive samples \cite{Friedrich2006, Doriese2016}. Moreover, they cover the entire tender X-ray range \cite{Uhlig2015} that is hardly accessible with both, grating and crystal spectrometers.

State-of-the-art TES- and MMC-based X-ray detectors achieve an energy resolution (FWHM) of $0.72\,\mathrm{eV}$ for $1.5\,\mathrm{keV}$ photons \cite{Lee2015} and of $1.25\,\mathrm{eV}$ for $5.9\,\mathrm{keV}$ photons \cite{Krantz2023} at near-unity quantum efficiency. However, despite of various ongoing developments, all targeting to improve the performance of these microcalorimeters, an energy resolution as low as of $100\,\mathrm{meV}$, required for investigating vibrations or $d$-$d$-excitations in soft X-ray spectroscopy or resonant inelastic X-ray scattering, has yet to be demonstrated. Against this background, we have recently proposed a novel type of microcalorimeter, called {\lSQ}, that promises to provide the required energy resolution \cite{Schuster2023_LSQ}.

\begin{SCfigure}
    \centering
    \includegraphics[width=.55\linewidth]{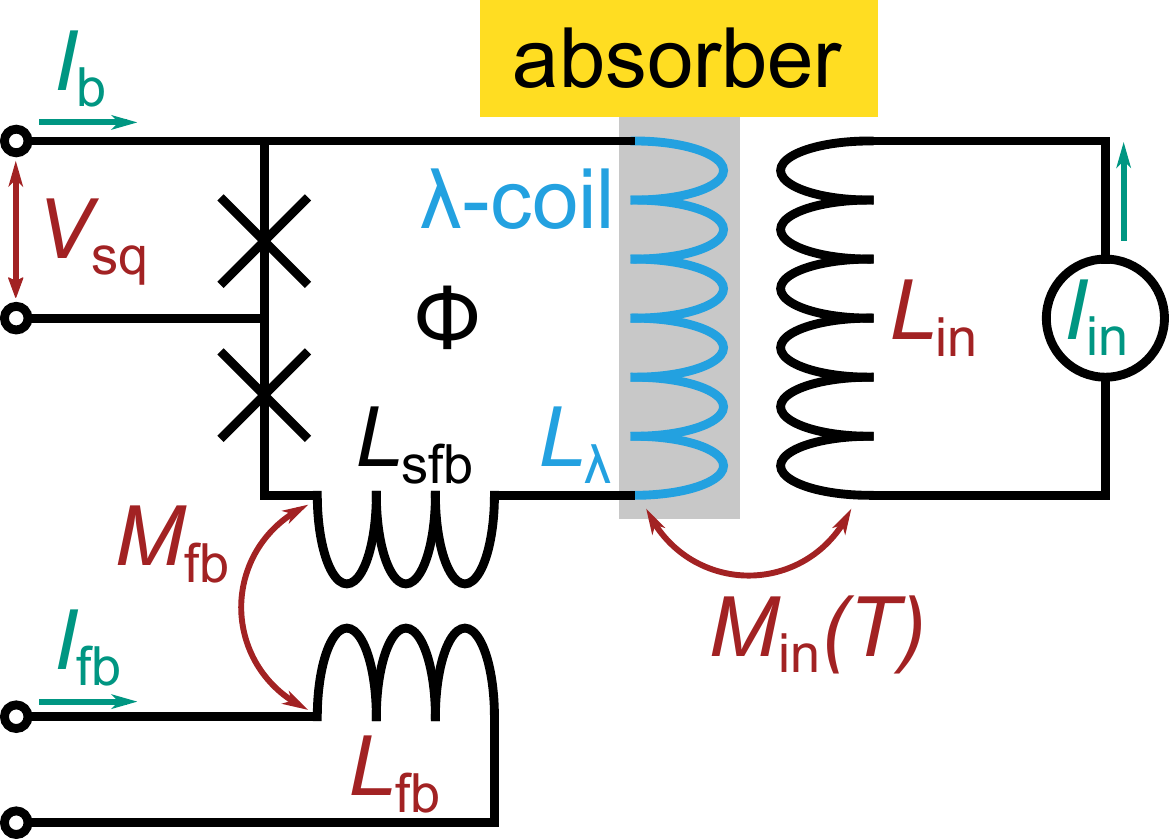}
    \caption{Simplified equivalent circuit diagram of a {\lSQ}. Inductors depicted in black are made from a superconductor with critical temperature $\Tc$. The {\lcoil} displayed in blue is made from a different superconductor with critical temperature $\Tl \ll \Tc$, and is in strong thermal contact with the absorber. The device is operated at temperature $T_0 \lessapprox \Tl$. A constant current $\Iin$ is injected into an input coil with inductance $\Lin$ and causes a temperature dependent magnetic flux within the SQUID loop via the a temperature dependent mutual inductance $\Min(T)$.}
  \label{fig:LSQ_shematic}
\end{SCfigure}

Fig.~\ref{fig:LSQ_shematic} shows a simplified equivalent circuit diagram of a \lSQ. Similar to a conventional dc-SQUID, it consists of a superconducting loop interrupted by two Josephson tunnel junctions, each with critical current $\Ic$, capacitance $\Cjj$ and normal state resistance $R$. The superconducting loop is divided into two parts, i.e. a section $\Lsfb$ that couples to a flux biasing coil and a section with inductance $\Ll$, denoted as {\lcoil}, which couples to an external input coil. Most parts of the device, including both, the input coil and flux biasing coil, the section $\Lsfb$, the Josephson junction electrodes, and the junction wiring, are made from a superconductor with a transition temperature $\Tc$ much larger than the device operating temperature $T_0$, i.e. $\Tc \gg T_0$. In contrast, the {\lcoil} is made from a different superconductor with a transition temperature $\Tl$ that barely exceeds the operating temperature, i.e. $\Tl \gtrapprox T_0$. As a consequence, the magnetic penetration depth $\lambda(T)$ of the {\lcoil} and hence the current distribution within the cross-section of the {\lcoil} shows a strong temperature dependence that affects both, the inductance $\Ll(T)$ of the {\lcoil} as well as the mutual inductance $\Min(T)$ between the {\lcoil} and the input coil. Assuming that a constant current $\Iin$ is injected into the input coil, the flux induced into the {\lSQ} then also becomes temperature sensitive: $\Phi(T) = \Iin \Min(T)$. As in a typical dc-SQUID, this change in flux can be measured as a change in voltage across or current through the {\lSQ}, depending on the mode of operation. By bringing a suitable particle absorber with specific heat $\Cabs$ into close thermal contact with the {\lcoil}, the temperature rise upon particle absorption can be accurately detected and measured.

\section{Design considerations of the {\lcoil}}

As the central sensing element, the {\lcoil} has an enormous influence on the performance of the {\lSQ}. Assuming a sophisticated readout chain in which the noise of subsequent amplifiers do not affect the noise performance of the {\lSQ}, e.g. by using an $N$-dc-SQUID series array as a first-stage low-temperature amplifier, the energy resolution $\dEFWHM$ of a {\lSQ} is set by two noise contributions, i.e. thermodynamic energy fluctuations $\SEtd$ caused by random energy fluctuations among the absorber, sensor and heat bath, and the noise contribution $\SEsq$ from the {\lSQ} itself. In the following, we will derive the conditions for the {\lcoil} which minimize the {\lSQ} noise contribution $\SEsq$.

\begin{figure}
    \centering
    \includegraphics[width=0.9\linewidth]{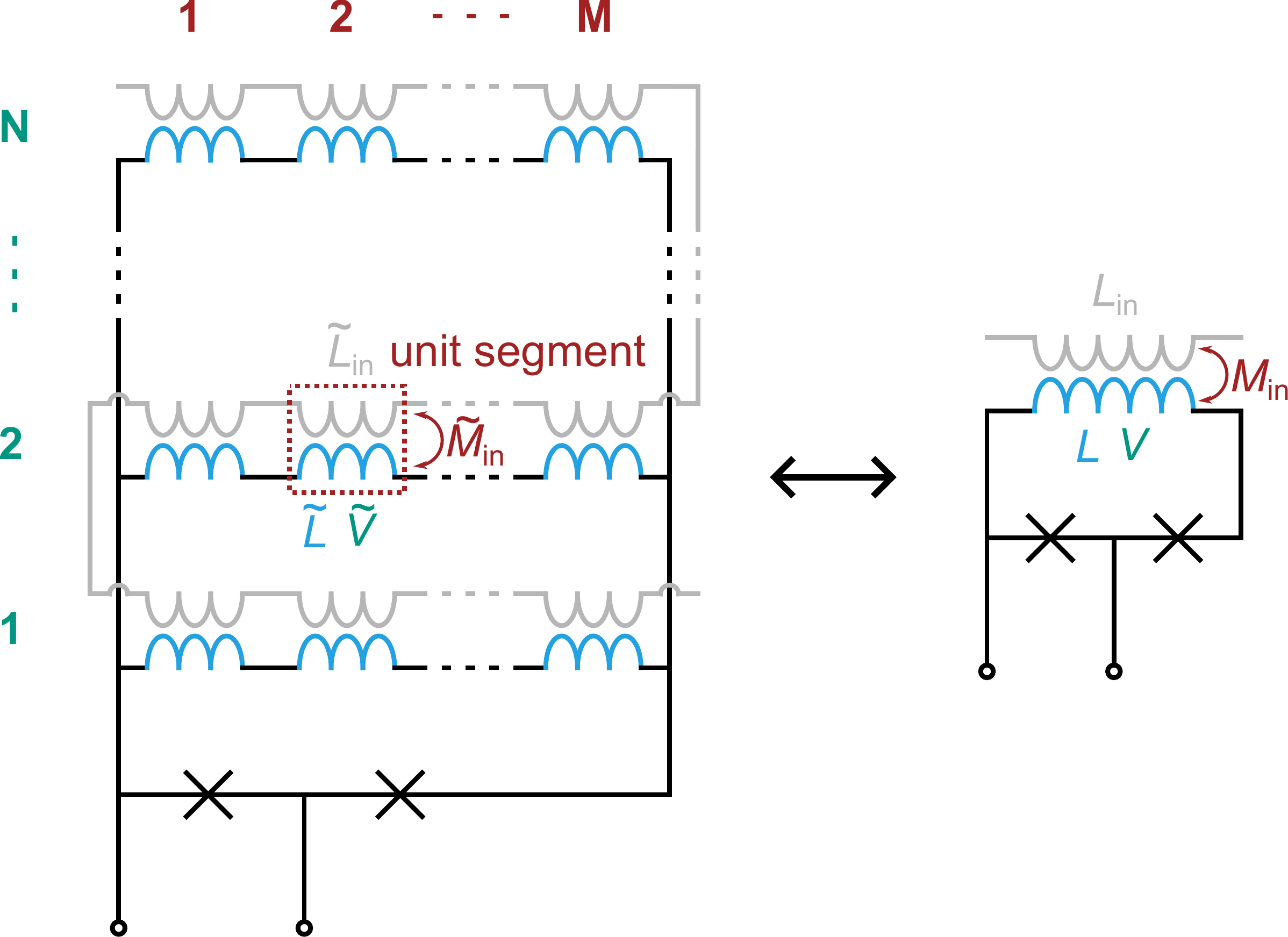}
    \caption{Schematic circuit diagram of a hypothetical {\lSQ}. The SQUID loop consists of $N$ identical segments connected in parallel, with each segment itself comprising $M$ unit elements connected in series. One such unit element is framed in red. Each unit element has an inductance $\tL$ with volume $\tV$, and couples to a unit of input loop with inductance $\tLin$ with a mutual inductance $\tMin$. The total arrangement is equivalent to a single effective {\lcoil} with inductance $L$, volume $V$ and mutual inductance $\Min$ to an input coil with inductance $\Lin$.}
    \label{fig:schematic}
\end{figure}

We assume a hypothetical {\lSQ} as schematically depicted in Fig.~\ref{fig:schematic}. For simplicity, the flux bias coil is neglected as it does not affect the temperature sensitivity. We consider a {\lcoil} entirely separable into small, identical unit elements, each with inductance $\tL$ and volume $\tV$. The loop comprises $N$ parallel rows of $M$ unit elements in series each, resulting in a total of $N \times M$ unit elements in total. Each unit element couples to a segment of input coil with inductance $\tLin$ via a mutual inductance $\tMin = \kappa \sqrt{\tL \tLin}$. Here, $\kappa$ is the geometric coupling factor. We can thus conclude the following relations for the total loop inductance $L$ and total volume $V$ of the {\lcoil} and the combined mutual inductance $\Min$ between the {\lSQ} and the input coil:
\begin{eqnarray}
    L &=& \frac{M}{N} \tL = \nu \tL, \label{eq:Lnu}\\
    V &=& M N \tV = \mu \tV, \label{eq:Vmu}\\
    \Min &=& M \tMin = \sqrt{\nu \mu} \tMin . \label{eq:Minmunu}
\end{eqnarray}
With only two free parameters to choose, we see immediately that we can not choose all three quantities independently. To quantify the remaining dependence, we have introduced the two parameters $\nu = M / N$ and $\mu = M N$ that fully define a specific layout of the {\lcoil}.

We use the geometric parameters $\nu$ and $\mu$ to express other relevant parameters: First, we assume that our {\lSQ} is optimized in terms of noise performance \cite{Tesche1977}, i.e. the SQUID screening parameter is $\betaL = 2 L \Ic / \Phi_0 = 1$ and the Stewart-McCumber parameter of the Josephson junctions is $\betaC = 2 \pi \Ic R^2 \Cjj / \Phi_0 = 1$. 
Assuming an established fabrication process for Josephson junctions, the critical current is set by the junction area $\Ajj$ and the critical current density $\jc$ via $\Ic = \Ajj \jc$. The junction capacitance consequently is $\Cjj = \Ajj \cjj = \Ic \cjj / \jc$, with the process-specific junction capacitance per unit area $\cjj$. From the conditions $\betaL, \betaC = 1$ we conclude:
 \begin{eqnarray}
    \Ic = \left[ \frac{\Phi_0}{2 \tL} \right]\frac{1}{\nu}, \label{eq:Icnu}\\
    R = \left[ \sqrt{\frac{2 \jc}{\pi \Phi_0 \cjj}} \tL \right]  \nu. \label{eq:Rnu}
\end{eqnarray}
In each expression, the term in brackets is independent of the specific design of the {\lSQ}, and depends only on fabrication parameters and the design of a unit element of the {\lcoil}.

The noise contribution $\SEsq$ caused by the {\lSQ}, expressed in fluctuations of the energy content, is given by the expression
\begin{equation}
    \SEsq = \Sphi \left( \frac{\partial \Phi}{\partial T} \frac{\partial T}{\partial E} \right)^{-2}.
\end{equation}
Here, $\Sphi$ denotes the magnetic flux noise of the SQUID. Moreover, $\frac{\partial \Phi}{\partial T}$ and \mbox{$\frac{\partial T}{\partial E} = 1 / \Ctot$} are the temperature-to-flux transfer coefficient of the {\lSQ} and the inverse total heat capacity of the detector, respectively. For an optimized dc-SQUID, the flux noise can be approximated by $\Sphi \approx 18 \kB T L^2 / R$ at the operating temperature $T$ \cite{SQUIDhandbook_Theory}. The total heat capacity of the detector comprises both, the absorber and the {\lcoil}, i.e. $\Ctot = \Cabs + \mu \tV c$, with $c$ the specific heat per unit volume of the {\lcoil}. The temperature-to-flux transfer coefficient $\frac{\partial \Phi}{\partial T}$ of the {\lSQ} is given by 
\begin{equation} \label{eq:dPhidT}
\frac{\partial \Phi}{\partial T}  = \Iin  \frac{\partial \Min}{\partial T} = \left[ \Iin 
  \frac{\partial \tMin}{\partial T} \right] \sqrt{\mu \nu },
\end{equation}
where, again, the term in brackets does not depend on the design of the {\lcoil} as a whole, but rather on the unit element.
Using Eq.~\ref{eq:Icnu} and Eq.~\ref{eq:Rnu} we obtain for the noise contribution
\begin{eqnarray}
    \SEsq &=& 18 \kB T \sqrt{\frac{\pi \Phi_0 \cjj}{2 \jc}} \tL \nu \left(  \left[ \Iin 
  \frac{\partial \tMin}{\partial T} \right] \sqrt{\mu \nu } \frac{1}{\Cabs + \mu \tV c} \right)^{-2} \\
  &=& \left[ 18 \kB T \sqrt{\frac{\pi \Phi_0 \cjj}{2 \jc}} \frac{\tL}{\Iin^2 \left( \frac{\partial \tMin}{\partial T} \right)^2} \right] \frac{\left( \Cabs + \mu \tV c\right)^2}{\mu} = \eta g\left( \mu \right). \label{eq:Sphi_final}
\end{eqnarray}
Here, we have introduced the substitutions
\begin{eqnarray}
    \eta &=& \left[ 18 \kB T \sqrt{\frac{\pi \Phi_0 \cjj}{2 \jc}} \frac{\tL}{\Iin^2 \left( \frac{\partial \tMin}{\partial T} \right)^2} \right] \quad \mathrm{and} \label{eq:eta_final} \\
   g\left( \mu \right) &=& \frac{\left( \Cabs + \mu \tV c\right)^2}{\mu}. \label{eq:gmu_final}
\end{eqnarray}
We see that $\SEsq$ neatly separates into a term $\eta$ which depends only on the unit element, but is independent of the specific layout of the {\lcoil}, and a function $g\left( \mu \right)$, which contains the dependence of the {\lSQ} noise $\SEsq$ on the arrangement of unit elements that comprises the {\lcoil}. It is interesting to note that only $\mu$ appears here, and that the parameter $\nu$ has dropped out. This can be understood as follows: An increase in $\nu$ causes a proportional increase in total inductance $L$ and a rise of \mbox{$\Min \propto \sqrt{\nu}$}. While the latter results in an increase of detector signal, the former affects the flux noise negatively. Ultimately, the signal-to-noise of the {\lSQ} remains unaltered by a change of the inductance $L$ by varying the parameter $\nu$ as introduced above.

To minimize the noise level $\SEsq$, we only have to consider the design parameter $\mu$. Since the prefactor $\eta$ has no influence on this optimization, we restrict our efforts to the function $g\left( \mu \right)$ and find
\begin{eqnarray}
   \frac{ \partial g\left( \mu \right)}{\partial \mu} &=& \frac{2 \left( \Cabs + \mu \tV c \right) c \tV }{\mu} -  \frac{\left( \Cabs + \mu \tV c \right)^2}{\mu^2}\label{eq:dgmu_dmu} \\
   \frac{ \partial^2 g\left( \mu \right)}{\partial \mu^2} &=&  \frac{2 \left( c \tV \right)^2}{\mu} -  \frac{4 \left( \Cabs + \mu \tV c \right) c \tV }{\mu^2}  + \frac{2 \left( \Cabs + \mu \tV c \right)^2}{\mu^3}  \label{eq:dgmu2_dmu2}
\end{eqnarray}
for its first and second derivative. In this way, we find that the function $g\left( \mu \right)$, and thus the {\lSQ} energy noise contribution $\SEsq$, has a minimum if the condition $\mu = \muopt$ is satisfied, with
\begin{equation} \label{eq:mumin}
    \muopt c \tV = \Csen = \Cabs.
\end{equation}
Thus, we can conclude that the layout of the {\lcoil} should be chosen such that the total specific heat of the {\lcoil} exactly equals the specific heat of the absorber $\Csen = \Cabs$. This resembles a well-known result for cryogenic microcalorimeters \cite{McCammon2005}.

\section{Conclusion}

The recently proposed {\lSQ} is a superconducting microcalorimeter with in-situ tunable gain and, with a proper choice of absorber and sensor material, promises to reach sub-eV energy resolution. In preparation for such a demonstration, we have presented theoretical design considerations related to the optimization of the layout of the {\lcoil}, which is the fundamental sensing element of a {\lSQ}. By sub-dividing the {\lcoil} into a large number of small unit elements, we could abstract the layout of any possible design of the {\lcoil} for a given fabrication method, and describe the remaining degrees of freedom in the design process by two parameters $\mu$ and $\nu$. When considering the influence of these parameters on the energy noise contribution $\SEsq$ of the {\lSQ}, we draw two conclusions. First, the total inductance $L$ of the {\lcoil} has no effect on the resulting energy noise, and can thus be chosen freely. Second, the noise contribution $\SEsq$ is minimized if the total volume of the {\lcoil} is chosen such that its specific heat directly equals that of the X-ray absorber: $\Csen = \Cabs$.

\bmhead{Acknowledgments}
C. Schuster acknowledges financial support by the Karlsruhe School of Elementary Particle and Astroparticle Physics: Science and Technology (KSETA).

\bibliography{99_bibliography.bib}

\end{document}